\documentclass[iop,apjl,tighten]{emulateapj}

\usepackage{natbib}
\slugcomment{To appear in The Astrophysical Journal}

\shorttitle{OXYGEN IN LOW- AND HIGH-ALPHA HALO STARS}
\shortauthors{RAM\'IREZ ET AL.}

\newcommand{\feh}{\mathrm{[Fe/H]}}
\newcommand{\teff}{T_\mathrm{eff}}
\newcommand{\logg}{\log g}

\newcommand{\oi}{O\,\textsc{i}}
\newcommand{\ofe}{\mathrm{[O/Fe]}}

\newcommand{\kms}{km\,s$^{-1}$}

\begin{document}

\title{OXYGEN ABUNDANCES IN LOW- AND HIGH-ALPHA FIELD HALO STARS AND \\
THE DISCOVERY OF TWO FIELD STARS BORN IN GLOBULAR CLUSTERS}

\author{I.\,Ram\'irez\altaffilmark{1},
        J.\,Mel\'endez\altaffilmark{2}, and
        J.\,Chanam\'e\altaffilmark{3}
}
\altaffiltext{1}{McDonald Observatory and Department of Astronomy,
                 University of Texas at Austin, 1 University Station, C1400
                 Austin, Texas 78712-0259, USA}
\altaffiltext{2}{Departamento de Astronomia do IAG/USP, 
                 Universidade de S\~ao Paulo,
                 Rua do M\~atao 1226, S\~ao Paulo, 05508-900, SP, Brasil}
\altaffiltext{3}{Departamento de Astronom\'ia y Astrof\'isica, 
                 Pontificia Universidad Cat\'olica de Chile,
                 Av.\ Vicu\~na Mackenna 4860, 782-0436 Macul, Santiago, Chile}

\begin{abstract}
\noindent Oxygen abundances of 67 dwarf stars in the metallicity range $-1.6<\feh<-0.4$ are derived from a non-LTE analysis of the 777\,nm \oi\ triplet lines. These stars have precise atmospheric parameters measured by Nissen and Schuster, who find that they separate into three groups based on their kinematics and $\alpha$-element (Mg, Si, Ca, Ti) abundances: thick-disk, high-$\alpha$ halo, and low-$\alpha$ halo. We find the oxygen abundance trends of thick-disk and high-$\alpha$ halo stars very similar. The low-$\alpha$ stars show a larger star-to-star scatter in [O/Fe] at a given [Fe/H] and have systematically lower oxygen abundances compared to the other two groups. Thus, we find the behavior of oxygen abundances in these groups of stars similar to that of the $\alpha$ elements. We use previously published oxygen abundance data of disk and very metal-poor halo stars to present an overall view ($-2.3<\feh<+0.3$) of oxygen abundance trends of stars in the solar neighborhood. Two field halo dwarf stars stand out in their O and Na abundances. Both G53-41 and G150-40 have very low oxygen and very high sodium abundances, which are key signatures of the abundance anomalies observed in globular cluster (GC) stars. Therefore, they are likely field halo stars born in GCs. If true, we estimate that at least $3\pm2$\,\% of the local field metal-poor star population was born in GCs.
\end{abstract}

\keywords{stars: abundances --- stars: Population II --- Galaxy: disk --- Galaxy: halo}

\section{INTRODUCTION}

Important clues to understand the formation and evolution of the Milky Way's halo and disk components, as well as any possible connections between them, are imprinted in the photospheric chemical composition of FGK-type dwarf stars. These objects have probably retained the chemical composition of the gas from which they formed, thus being excellent tracers of Galactic chemical evolution (GCE). Moreover, their long lifetimes, in particular for the G and K types, allow us to probe the Milky Way's GCE over many billions of years. Consequently, combined with kinematics and information on stellar ages, chemical composition analyses of FGK dwarf stars have the potential to be useful for reconstructing the history of our Galaxy.

The simplest picture for the Galactic halo formation involves a monolithic collapse which leads to a halo star population showing a strong correlation between orbital eccentricity an overall metal abundance \citep{eggen62}. Although historically important, this model has long been known to be incomplete. Metallicity determinations of giant stars in globular clusters, for example, led \cite{searle78} to conclude that the halo was formed in a more ``chaotic'' fashion. Indeed, state-of-the-art simulations show that the properties of the stellar component of galactic halos could be heavily influenced by merging events that occur as the galaxy assembles \cite[e.g.,][]{abadi03,guo08,read08,stewart08,scannapieco09}. In these scenarios, rather than consisting of a single simple evolving population, the halo is expected to contain sub-structures as remnants of its formation history. The discovery and detailed characterization of halo streams and tidal debris heavily support this idea \cite[e.g.,][]{majewski93,helmi08,klement10,majewski12}.

Very strong evidence for halo sub-structures in the solar neighborhood has been recently provided in a series of papers by \cite{nissen10,nissen11}, and \cite{schuster12}. From a detailed spectroscopic analysis of 94 dwarf stars in the $\feh$ range from $-1.6$ to $-0.4,$\footnote{In this work we use the standard definitions: $\mathrm{[X/Y]}=\log(N_\mathrm{X}/N_\mathrm{Y})-\log(N_\mathrm{X}/N_\mathrm{Y})_\odot$, and $A_\mathrm{X}=\log(N_\mathrm{X}/N_\mathrm{H})+12$, where $N_\mathrm{X}$ is the number density of element X.} \citet[][hereafter NS10]{nissen10} found that stars with halo kinematics separate into two groups based on their $\alpha$-element abundances (in their case quantified by the average abundance of Mg, Si, Ca, and Ti). NS10 argue that the halo stars in the low-$\alpha$ group could have been accreted from satellite galaxies, possibly $\omega$\,Cen. Their abundance analysis of heavier elements, particularly Na and Ba/Y, however, showed that the $\omega$\,Cen and low-$\alpha$ halo star connection is weak, unless chemical evolution within the satellite galaxy was different for its inner and outer regions, an idea supported by the observation of an abundance gradient in a dwarf galaxy \cite[cf.][]{nissen11}. Finally, \cite{schuster12} show that the low-$\alpha$ halo stars are about 2--3\,Gyr younger than the high-$\alpha$ halo stars and that these two groups exhibit different orbital properties, with the low-$\alpha$ stars having very eccentric orbits, larger $r_\mathrm{max}$ (maximum distance from the Galactic center), and larger $z_\mathrm{max}$ (maximum distance from the Galactic disk).

A very important chemical element missing from the NS10 paper series is oxygen. As the third most abundant element in the universe and in stellar atmospheres, after H and He, and having one of the best identified production sites of all elements as well as reliable supernovae yields, oxygen is crucial for GCE studies. Furthermore, oxygen is a key element in the investigation of abundance variations in globular clusters (GCs). Stars with enhanced Na are known to be depleted in O, i.e., they follow the well-known oxygen-sodium anti-correlation in GCs \cite[e.g.,][]{gratton04,cohen05,yong05,alves12}. A number of recent studies have investigated the contribution of GCs to the build-up of the field halo population \cite[][]{yong08,carretta10,martell10,martell11}, but none of them have found field stars with both high Na and low O abundances. Since NS10 have already studied Na in their sample of metal-poor stars, the addition of oxygen allows us to assess to which level the halo field has been contaminated by stars formed in GCs.

Determining reliable oxygen abundances in metal-poor dwarf stars is not a straightforward task. Few spectral features due to oxygen are available in the visible spectrum and all are affected by a number of model uncertainties or severe line blending. Our past experience successfully deriving oxygen abundances from a restricted non-LTE analysis of the 777\,nm \oi\ triplet \citep[][hereafter R07]{ramirez07} now allows us to infer them. In this work, we derive oxygen abundances for as many as possible of the stars in the NS10 study in order to better understand the nature of the two distinct halo populations in the solar neighborhood.

\section{SAMPLE AND SPECTROSCOPIC DATA}

Given the careful sample selection and high precision of the stellar parameters and elemental abundances derived by NS10, we adopted their sample in our work, as stated above. We collected high-quality spectra of the 777\,nm region for as many of these stars as possible. We started by searching for data in our own spectral libraries and then in publicly available data archives. We complemented this data set with new observations, as described below.

First, we used data from the R07 work on oxygen abundances in nearby stars. Nearly all of these spectra were acquired with the R.\,G.\,Tull coud\'e spectrograph on the 2.7\,m Telescope at McDonald Observatory, and reduced in the standard manner using IRAF,\footnote{IRAF is distributed by the National Optical Astronomy Observatories, which are operated by the Association of Universities for Research in Astronomy, Inc., under cooperative agreement with the National Science Foundation -- http://iraf.noao.edu} as described in R07. One spectrum from the High Resolution Spectrograph (HRS) on the Hobby-Eberly Telescope (HET), reduced also as in R07, and one spectrum from the VLT-UVES POP (Paranal Observatory Project) library \citep{bagnulo03} were also used. The McD-Tull data have a spectral resolution $R=\lambda/\Delta\lambda\simeq60,000$ while the HET-HRS and VLT-UVES spectra have $R\simeq120\,000$ and $R\simeq80\,000$, respectively.

Then, we searched for spectra taken with the HIRES spectrograph \citep{vogt94} at the Keck\,I Telescope. We found 11 stars available in the Keck Observatory data archive covering the 777\,nm \oi\ triplet. These spectra were reduced using MAKEE,\footnote{http://spider.ipac.caltech.edu/staff/tab/makee} a data reduction tool developed by T.\,A.\,Barlow specifically for reduction of Keck-HIRES spectra. In some cases, we re-reduced the archive spectra with MAKEE by fine-tuning the extraction parameters in order to improve the results. Most of the HIRES spectra have $R\simeq67\,000$, but some of them were taken at $R\simeq50\,000$ or $R\simeq100\,000$.

We found 27 of the NS10 stars in the R07 work and 11 more in the Keck-HIRES archive. In order to build a more statistically significant sample, we performed spectroscopic observations of 24 additional stars using the MIKE spectrograph on the 6.5\,m Magellan/Clay Telescope at Las Campanas Observatory. We acquired the data for these stars in four observing runs in July, September, and November of 2011, as well as in February of 2012. We used the MIKE standard setting with the narrowest slit (0.35\,arcsec width), which provides complete wavelength coverage from 3350 to 9500\,\AA, including the 777\,nm oxygen triplet, at a spectral resolution $R=65\,000$ in the oxygen triplet region \citep{bernstein03}. The signal-to-noise ratios of these spectra vary from star to star from about 100 to 500 at 777\,nm. The MIKE spectra were reduced using the CarnegiePython pipeline.\footnote{http://obs.carnegiescience.edu/Code/mike} One more star (G150-40) was observed at McDonald Observatory in April of 2012. Its spectrum ($R\simeq60\,000$, $S/N\simeq150$) was reduced as in R07.

The high quality of our spectroscopic data allows a very precise measurement of the equivalent widths (EWs) of the three lines of the oxygen triplet. Based on the spectral resolution and typical signal-to-noise ratio of most of our data ($R\simeq60\,000$, $S/N\simeq200$), we estimate an error of about 0.7\,m\AA\ for each EW measurement (using \citealt{cayrel88} formula). The actual EW error is likely larger due to uncertainties in the continuum determination, but tests made measuring the EW of a number of lines while varying slightly the continuum location showed that these errors are unlikely to be larger than about 1.0\,m\AA. Thus, we estimate our EW errors to be about 1.2\,m\AA.

Three of the stars from the NS10 work, and for which we did not find spectra or were unable to acquire them, have precise (errors of order 2\,m\AA) EWs of the oxygen triplet by \cite{akerman04}. We adopted those EW values in our work. Also, for G75-31, although a spectrum from R07 is available, it is of relatively low quality, and therefore we preferred to use the EW values for the star measured by \cite{nissen02} using a higher quality VLT-UVES spectrum. For the Sun, which we use to transform the abundances from absolute ($A_\mathrm{O}$) to relative ([O/H]), we adopted the solar EWs by R07. These values were obtained as the average of EWs measured in three solar spectra, two skylight observations and one asteroid reflected sunlight spectrum, which were shown to be in very good agreement.

\section{OXYGEN ABUNDANCES} \label{s:oxygen}

The EWs of the three lines of the \oi\ triplet at 777\,nm were measured by fitting Voigt profiles using IRAF's splot task, except for weak lines with relatively low signal-to-noise ratio spectra. The Gaussian profile is a very good approximation to the real line shapes considering the spectral resolution of most of our data and the fact that these lines are not so strong in this metallicity regime. Voigt profile fits are more accurate for strong lines with extended wings, but they tend to confuse noise with wing depth for weak lines in low S/N spectra. In those cases a Gaussian fit often works better. Our measured equivalent widths were then used to derive the oxygen abundances using a standard curve-of-growth (COG) approach, as described below.

The abfind driver of the 2010 version of the spectrum synthesis code MOOG \citep{sneden73}\footnote{http://www.as.utexas.edu/~chris/moog.html} was used to compute COGs and derive the oxygen abundances from our EW measurements. We used the new MARCS grid of model atmospheres \citep{gustafsson08}. The stellar parameters $\teff$, $\logg$, and $\feh$ adopted are those derived by NS10. For a number of stars (6), NS10 provide two sets of stellar parameters, each derived from a different spectrum. For all these stars, the two sets of parameters are in excellent agreement considering the observational errors. In these cases, we adopted the averages of the two sets of atmospheric parameters derived by NS10.

NS10 estimate their differential errors for $\teff$, $\logg$, and $\feh$ at 30\,K, 0.05\,dex, and 0.03\,dex, respectively. Using the reference stars HD\,22879 and HD\,76932, we can obtain an estimate for the oxygen abundance ([O/H]) error by calculating the variation of their absolute oxygen abundances, $A_\mathrm{O}$, for a given change in stellar parameters. This procedure is safe because these uncertainties are only weakly correlated. Propagated into our abundance analysis, the errors in stellar parameters translate into a 0.025\,dex uncertainty for [O/H]. The microturbulent velocities, $v_t$, were also adopted from NS10, where no estimate of the $v_t$ error is given. If we assume a $v_t$ error of 0.2\,\kms, which is a conservative estimate considering the high precision of the strictly differential work by NS10, the uncertainty introduced to the oxygen abundance is only about 0.015\,dex. The EW uncertainty of each of the triplet lines translates into a 0.020\,dex error, which implies a total oxygen abundance error of about 0.035\,dex (internal error only). Model uncertainties and noise or instrumental defects are probably responsible for the larger line-to-line scatter (in the oxygen abundances inferred for the three of the triplet lines) seen in some of our sample stars.

The spectrum synthesis code used to derive our oxygen abundances assumes local thermodynamic equilibrium (LTE). However, it is well known that in the photospheres of cool stars, both dwarfs and giants, the \oi\ 777\,nm triplet spectral feature is formed under conditions far from the LTE approximation. Using a well-justified two-level approximation, \cite{kiselman93} elegantly demonstrates that the non-LTE effect is due to an infrared mean intensity that departs from its LTE value (the Planck function) at depths where the triplet lines are formed \cite[see also][]{eriksson79,kiselman01}. Since the radiation field is stronger in warmer stars and the gas densities are smaller in lower surface gravity stars, the errors due to non-LTE effects are larger for warmer $\teff$ and lower $\logg$ values. An important $\feh$-dependence is also anticipated, with larger non-LTE effects at lower [Fe/H] due to a decrease of continuum opacity and smaller rate of collisions \cite[see, e.g.,][]{fabbian09}. In the solar case, the non-LTE correction to the oxygen abundances inferred from the triplet is about 0.2\,dex \cite[e.g.,][]{kiselman93,takeda94,ramirez07,fabbian09}. In warmer stars ($\teff\simeq6250$\,K), these corrections can be as high as $\sim0.4$\,dex, which means that even in a solar relative analysis, differential non-LTE errors of $\sim0.2$\,dex could be introduced. Thus, non-LTE corrections to the LTE abundances derived with MOOG must be applied.

A number of authors have computed non-LTE corrections for FGK-type stars of several metallicities, including those of our targets. There is good qualitative agreement between the various calculations available in the literature, but differences of order 0.1\,dex between them are common. In this work, we use the non-LTE corrections tabulated by R07. The main reason for this choice is that the same corrections were applied to the abundances of disk stars which we will use later in Section~\ref{s:discussion}, and similar corrections were applied to the very metal-poor stars from the \cite[][hereafter M06]{melendez06} work, which we will also use in Section~\ref{s:discussion}. To prevent systematic biases arising from the use of very different non-LTE prescriptions in our final analysis, we decided to make a choice that will bring the non-LTE oxygen abundances to approximately the same scale.

R07 computed their non-LTE corrections using an oxygen model atom with 54 levels and 242 transitions by \cite{allende03a}, with a few minor improvements. \cite{kurucz93:cd13} atmosphere models were employed to determine the level populations by solving the rate equations with TLUSTY \citep{hubeny88}. Although this implies a certain level of inconsistency within this work, because the model atmospheres used to derive the LTE oxygen abundances are from the MARCS grid, note that the non-LTE corrections were derived with respect to LTE abundances computed also from \cite{kurucz93:cd13} models. It is expected that the non-LTE corrections are less sensitive to the choice of model atmosphere grid than the absolute values of the abundances, either LTE or non-LTE. Indeed, \cite{fabbian09} have shown that the model-dependence in this context is only important for stars with $\feh<-2.5$, i.e., stars with metallicities below that of our most metal-poor sample star. Once the non-LTE level populations were computed with TLUSTY, spectrum synthesis was performed using SYNSPEC \citep{hubeny95}.

As acknowledged by R07, one of the deficiencies of their non-LTE calculations is the neglect of inelastic collisions with neutral H, which tend to bring the level populations closer to their LTE values. R07 noticed that their non-LTE oxygen abundances presented systematic offsets between the three lines of the triplet, and suggested empirically correcting for these offsets to reduce the line-to-line scatter. The nature of these offsets is qualitatively well explained by the fact that inelastic collisions with neutral H were ignored, but detailed calculations were not made to confirm this hypothesis. However, if true, these empirical corrections are roughly taking this effect into account.

The importance of inelastic collisions with neutral H in non-LTE calculations is typically parameterized by the multiplicative factor $S_\mathrm{H}$ to the \cite{drawin68} formula, as suggested by \cite{steenbock84}. Interestingly, the improved non-LTE computations by \cite{fabbian09}, including the impact of collisions with neutral H, show that their non-LTE corrections to the solar oxygen abundance with $S_\mathrm{H}=1$ are, albeit fortuitously, in good agreement with those by R07. In their three-dimensional hydrodynamic analysis of oxygen line formation in the solar photosphere, \cite{pereira09a,pereira09b} find that $S_\mathrm{H}\simeq1$ provides an excellent fit to the observational data. Note, however, that \cite{ramirez06} find that a one-dimensional static model atmosphere spectrum with $S_\mathrm{H}=10$ reproduces the 777\,nm \oi\ triplet line profile better than one with $S_\mathrm{H}=1$ in the case of the moderately metal-poor star BD\,+17\,4708, although the derived oxygen abundance in that case appears too high.

We measured oxygen abundances, $A_\mathrm{O}$, for each of the three triplet lines. The solar values were used to convert these quantities into [O/H] by averaging the three relative abundances. The same solar-differential line-by-line procedure was employed to derive the LTE and non-LTE relative abundances. Finally, [O/Fe] values were determined using the very precise ($\simeq0.03$\,dex error) iron abundances inferred by NS10: [O/Fe]=[O/H]--[Fe/H]. The mean error in the non-LTE [O/Fe] abundance ratios derived in this work is 0.05\,dex.

Our derived oxygen abundances, both in LTE and non-LTE, as well as the stellar parameters by NS10, their $\alpha$/Fe classification, and the source of our spectroscopic data for oxygen abundance analysis, are given in Table~\ref{t:ofe}.

\section{DISCUSSION} \label{s:discussion}

\subsection{Oxygen in Low- and High-$\alpha$ Halo Stars}

\begin{figure}
\centering
\includegraphics[bb=70 365 450 1027,width=8.85cm]
{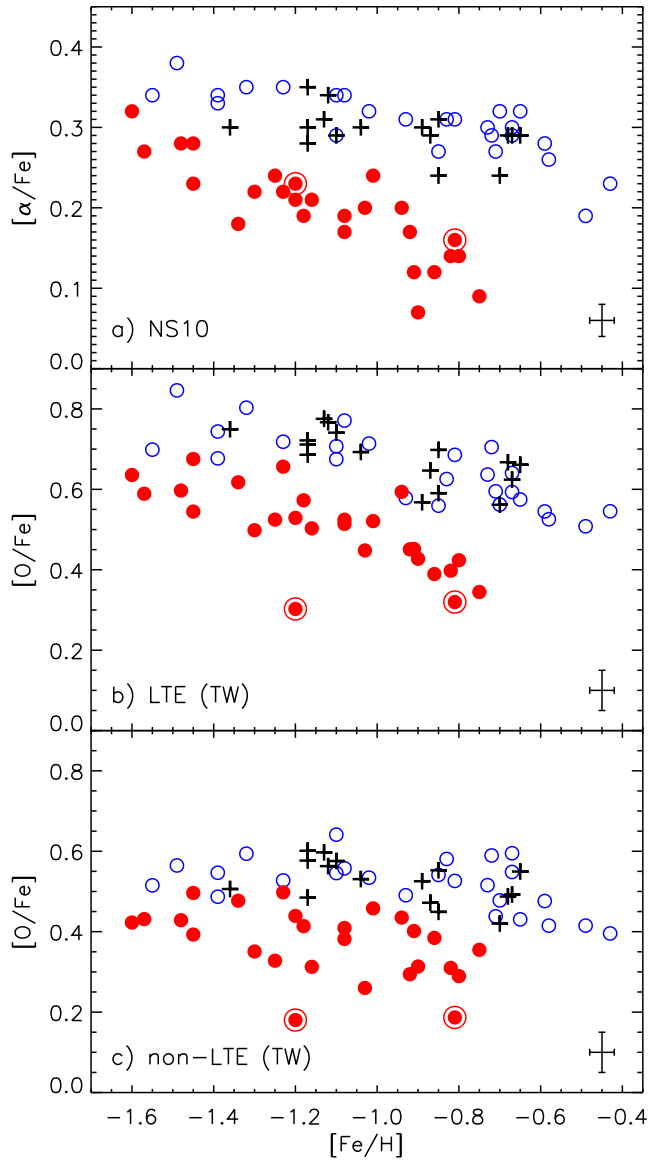}
\caption{{\bf a)} $\alpha$-element abundances as a function of $\feh$ for the thick-disk (crosses), high-$\alpha$ (open circles), and low-$\alpha$ (filled circles) halo stars from NS10 included in this work. The two low-$\alpha$ stars with large open circles surrounding them are G53-41 ($\feh=-1.20$) and G150-40 ($\feh=-0.81$). {\bf b)} LTE oxygen abundances as a function of $\feh$ for the stars plotted in panel a). {\bf c)} Non-LTE corrected oxygen abundances. The $\alpha$-element abundances are those derived by NS10; the oxygen abundances are those derived in this work (TW). Typical error bars are shown at the bottom right side of each panel.}
\label{f:ofe}
\end{figure}

NS10 convincingly showed that halo stars in the solar neighborhood divide into two groups based on their $\alpha$-element abundances. Their main result is reproduced here in Figure~\ref{f:ofe}a, but note that we show only the 67 stars for which we have also derived oxygen abundances. The NS10 work is based on the analysis of 94 stars. Nevertheless, our sub-sample is large enough to show the differences and similarities between the three groups plotted in Figure~\ref{f:ofe}a. NS10 sorted their stars into high-$\alpha$ halo stars (open circles in Figure~\ref{f:ofe}), low-$\alpha$ halo stars (filled circles), and thick-disk members (crosses). The latter were intentionally included in the work by NS10 for comparison with the old disk population, and they were disentangled from the halo group by their total Galactic space velocity, with thick-disk stars being slower than 175\,\kms. While the high-$\alpha$ and thick-disk groups appear to have indistinguishable [$\alpha$/Fe] versus $\feh$ trends, the low-$\alpha$ group, as their name suggests, have systematically lower [$\alpha$/Fe] at any given $\feh$. Note, however, that the low and high-$\alpha$ groups appear to merge at the lowest $\feh$ values covered by this sample. Although oxygen is an $\alpha$ element, it was not included in the NS10 work; their [$\alpha$/Fe] values are based on Mg, Si, Ca, and Ti abundances.

The LTE oxygen abundance patterns we derive for the NS10 stars are plotted in Figure~\ref{f:ofe}b. The exact same behavior is noted for the three groups, i.e., low-$\alpha$ stars have also low oxygen abundances. Note that the [O/Fe] abundance ratios appear to increase with lower $\feh$ in a relatively rapid manner, similar to the [$\alpha$/Fe] case. However, this trend is mostly due to non-LTE effects, which become important at low [Fe/H] and the somewhat warmer $\teff$ values of the more metal-poor stars in this sample (cf.~Section~\ref{s:oxygen}). As shown in Figure~\ref{f:ofe}c, the [O/Fe] versus [Fe/H] relation, corrected for non-LTE effects, is nearly flat up to $\feh\simeq-0.7$ for both the low- and high-$\alpha$ groups, and also for the thick disk. A hint of a knee towards lower [O/Fe] abundance ratios at $\feh\simeq-0.7$ for the high-$\alpha$ group is detected, and it will be confirmed later in this paper when we introduce additional (literature) data. Hereafter, the oxygen abundances used in our discussion are those corrected for non-LTE effects, i.e., those shown in Figure~\ref{f:ofe}c.

The star-to-star scatter in [O/Fe] at a given [Fe/H] appears to be the largest for the low-$\alpha$ group. Simple linear fits to the [O/Fe] versus [Fe/H] relation result in a 1\,$\sigma$ scatter of 0.056\,dex for the high-$\alpha$ stars, 0.047\,dex for the thick-disk stars, and 0.075\,dex for the low-$\alpha$ group. We note, however, that there are two stars which show very low oxygen abundances ([O/Fe]$<0.2$) and they increase significantly the scatter of the low-$\alpha$ group. Excluding these stars, the 1\,$\sigma$ scatter of the low-$\alpha$ group (0.058\,dex) is essentially the same as that of the high-$\alpha$ halo stars.

As pointed out by NS10, and as mentioned before, the low- and high-$\alpha$ populations appear to merge below $\feh\simeq-1.4$, making it more difficult to disentangle them using chemical abundances. Note, however, that the separation is more clear if $[\alpha/\mathrm{Fe}]$ is used instead of, for example, [Mg/Fe]. Moreover, if we calculate a new $\alpha$ element abundance as $[\alpha/\mathrm{Fe}]'=(4[\alpha/\mathrm{Fe}]+\ofe)/5$, the separation appears even clearer.

Our sample includes five low-$\alpha$ and two high-$\alpha$ halo stars with $\feh<-1.4$. With the exception of one low-$\alpha$ star, which has an [O/Fe] abundance ratio similar to the average of that for the high-$\alpha$ stars at $\feh<-1.4$ (HD\,219617, $\feh=-1.45$, $\ofe=0.50$), Figure~\ref{f:ofe}c strengthens the classification suggested by NS10. Although four of the five low-$\alpha$ halo stars with $\feh<-1.4$ have [O/Fe] abundance ratios that are marginally consistent, considering the 1-$\sigma$ error, with the average [O/Fe] of the most metal-poor high-$\alpha$ stars, it is highly unlikely that all, simultaneously, have underestimated oxygen abundances. Indeed, the probability that all these four stars have [O/Fe] abundance ratios greater by $1\,\sigma$ is only about 1\,\%. Similarly, the probability that the two high-$\alpha$ stars both have [O/Fe] lower by $1\,\sigma$ is about 10\,\%. Thus, the ambiguity regarding population membership at $\feh<-1.4$ may affect only one star: HD\,219617. If instead of considering this object as a low-$\alpha$ star we assume that it belongs to the high-$\alpha$ group, our conclusions remain unaltered. For example, the $1\,\sigma$ scatter values mentioned in the previous paragraph do not change by more than 0.001\,dex. Therefore, we conclude that the ``merging'' of low- and high-$\alpha$ populations at $\feh<-1.4$ do not affect our conclusions in a significant manner.

The two stars with the lowest oxygen abundances in Figure~\ref{f:ofe}c are G53-41 ($\feh=-1.20$, $\ofe=0.18$) and G150-40 ($\feh=-0.81$, $\ofe=0.19$). Interestingly, these stars have perfectly normal low-$\alpha$ abundances, i.e., although their [$\alpha$/Fe] abundance ratios are low compared to the high-$\alpha$ and thick-disk stars, they are not significantly lower than those of a typical low-$\alpha$ halo star. NS10 noticed that these stars also have very high Na abundances, suggesting that the gas which formed them was polluted by nucleosynthesis products from nearby asymptotic giant branch (AGB) stars, similarly to Na-enhanced stars in globular clusters (see Section~\ref{s:gc} for more details).

\begin{figure}
\centering
\includegraphics[bb=85 365 450 628,width=8.8cm]
{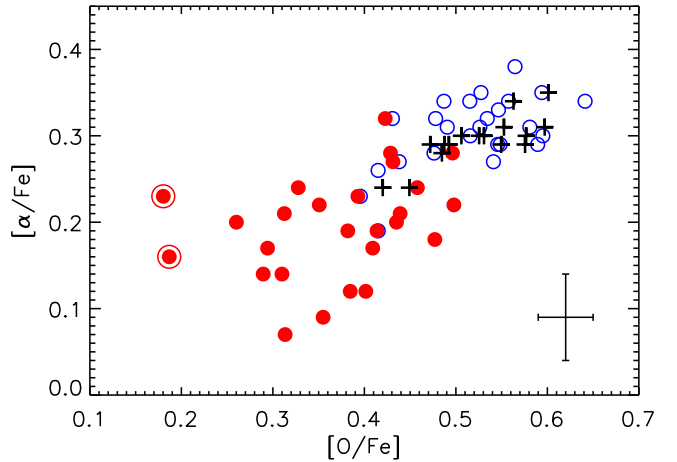}
\caption{[$\alpha$/Fe] versus [O/Fe] relation for the stars in Figure~\ref{f:ofe}. Typical error bars are shown at the bottom right corner.}
\label{f:ao}
\end{figure}

Not surprisingly, the oxygen and $\alpha$-element abundances correlate well, as shown in Figure~\ref{f:ao}, with the exception of the two stars with the lowest [O/Fe] abundance ratios mentioned above. Although the star-to-star scatter of the [$\alpha$/Fe] versus [O/Fe] relation is low for the high-$\alpha$ and thick-disk stars (1\,$\sigma$ scatter of 0.033 and 0.017\,dex, respectively, for a simple linear fit), that for the low-$\alpha$ stars is clearly larger (0.057\,dex), even if we exclude the two stars with [O/Fe]$<0.2$ (0.055\,dex).

\begin{figure*}
\centering
\includegraphics[bb=70 370 730 630,width=16.0cm]
{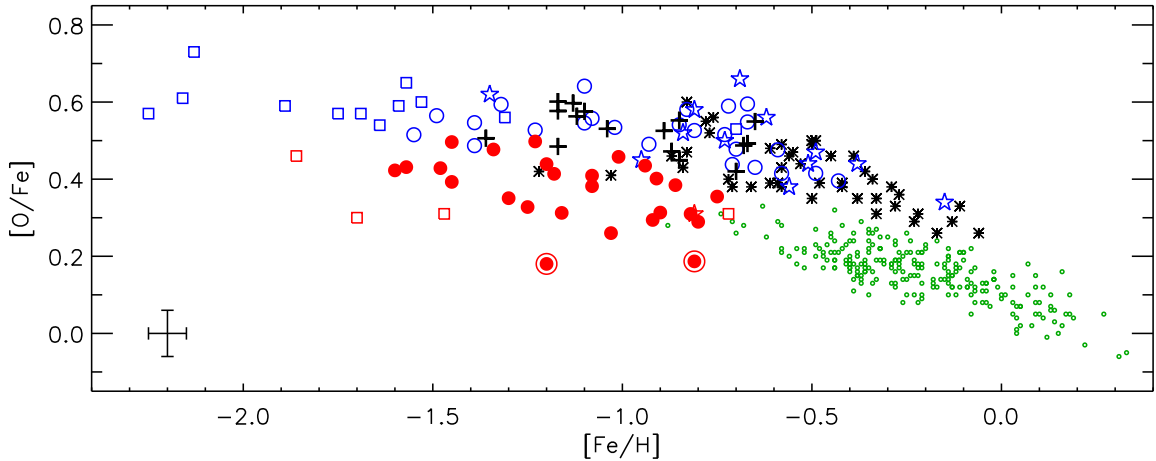}
\caption{Non-LTE oxygen abundances inferred from the 777\,nm \oi\ triplet as a function of $\feh$. Crosses, large open circles, and filled circles correspond to the stars plotted in Figure~\ref{f:ofe}. Small open circles and asterisks represent thin- and thick-disk stars, respectively, from R07. The membership criterion for the disk stars is based on the abundances, not the kinematics. Hence the perfect separation of thin- and thick-disk stars. Open squares and star symbols correspond to halo stars from M06 and R07, respectively. Typical error bars are shown at the bottom left corner.}
\label{f:ofe_all}
\end{figure*}

\subsection{Galactic Chemical Evolution of Oxygen: 777\,nm \oi\ Triplet Analyses of Solar Neighborhood Stars}

In order to put the oxygen abundance trends of low- and high-$\alpha$ halo stars, as inferred from the 777\,nm \oi\ triplet lines, into a broader context, in Figure~\ref{f:ofe_all} we show also the non-LTE corrected [O/Fe] abundance ratios by M06 and R07. Stars from the NS10 work included in these previous studies were excluded; i.e., we used the results obtained in this paper instead of the literature values. The use of M06 and R07 data allows us to study the GCE of oxygen from $\feh=-2.3$ to +0.3, and therefore that of the different stellar populations it includes, albeit only their solar neighborhood members. We note that, although not identical, the stellar parameter determination and non-LTE corrections applied in M06, R07, NS10, and this work, are similar. This prevents systematic offsets from biasing our data and artificially introducing noise to the chemical abundance trends. As shown in Figure~\ref{f:ofe_all}, stars of common populations but from different data sets connect nicely, suggesting that systematic differences between these three studies are indeed not very important.

The stars from M06 are shown with open squares in Figure~\ref{f:ofe_all}; they are all metal-poor main-sequence and turn-off stars with halo kinematics selected by \cite{akerman04} and \cite{nissen04}. These authors mention that their sample has halo kinematics, without given further details. M06 verified that they indeed have halo kinematics based on their large total Galactic space velocities ($V_\mathrm{tot}$). In fact, the Galactic space velocities of this sample fulfill the same criteria used by NS10 for selecting halo stars, meaning that they all have $V_\mathrm{tot}$ larger than 180 \kms, except one star, BD\,+08\,3095, which has halo kinematics according to its admittedly uncertain {\it Hipparcos} parallax, but it may actually be a thick-disk star based on its spectroscopic parallax. Nevertheless, the uncertain membership of this single star does not affect our results.

Most of the objects from M06 appear to be the natural extension of the high-$\alpha$, high-oxygen abundance population down to $\feh=-2.3$. Four objects from M06, however, seem to have low oxygen abundance and are more likely associated with the low-$\alpha$ population. These stars are CD\,$-$42\,14278 ($\feh=-1.86$, $\ofe=0.46$), G24-3 ($\feh=-1.47$, $\ofe=0.31$), HD\,146296 ($\feh=-0.72$, $\ofe=0.31$), and HD\,160617 ($\feh=-1.70$, $\ofe=0.30$). They can be identified in the color version of Figure~\ref{f:ofe_all} as red open squares.

The R07 data includes thin-disk, thick-disk, and a few halo stars. Along with the M06 and NS10 stars, objects from R07 are all in the solar neighborhood, i.e., within a volume with radius of a few hundreds of parsecs. This region is dominated by thin-disk stars. Studying thick-disk or halo stars {\it in-situ} would require observing objects at distances of order 1\,kpc in the direction perpendicular to the Galactic plane, as the thick-disk scale height has been claimed to be between 0.6 and 1.5\,kpc \cite[e.g.,][]{gilmore83,siegel02,cabrera05,dejong10,mateu11}. Although a kinematic criterion could be applied to disentangle the disk populations in the solar neighborhood, it has been shown that there is not a perfect one-to-one correspondence with the abundances (see also below). Thus, for local stars within a few hundreds of parsecs of distance, a chemical tagging approach would be more appropriate. In this work, we separate thin-disk and thick-disk stars by their [O/Fe] abundance ratios instead of their kinematics. A broken line with nodes at $(\feh,\ofe)=(-0.03,0.22),(-0.55,0.35),(-1.00,0.35)$ is (somewhat arbitrarily) used as the thin/thick disk boundary. In Figure~\ref{f:ofe_all}, our chemically-tagged thin-disk stars are shown with small open circles, while the thick-disk members are plotted with bold asterisks. The latter also appear to be a natural extension of the NS10 data for thick-disk stars up to $\feh\simeq-0.1$. The apparent lack of disk stars with [O/Fe] abundance ratios intermediate between those of a typical thin-disk and thick-disk star at $\feh\simeq-0.5$ is most likely due to sample selection biases (see Section~\ref{s:kinematics}).

A few objects with halo kinematics from R07 are shown in Figure~\ref{f:ofe_all} with five-pointed stars. These stars have a probability greater than 50\,\% of being halo members according to the kinematic criterion employed by R07 (their Sect.~3.3). In summary, the thin disk, thick disk, and halo populations are assumed to have Gaussian Galactic space velocity distributions, with mean $U,V,W$ values and velocity dispersions given by \cite{soubiran03} for the thin/thick disk and \cite{chiba00} for the halo.

Only one of the halo stars from R07 seems to belong to the low-$\alpha$ population: HIP\,4544 ($\feh=-0.81$, $\ofe=0.30$). We note that, according to R07, HIP\,4544 has a high probability of being a thick-disk member (43\,\%), but its low oxygen abundance clearly suggests that it is a low-$\alpha$ halo object instead. \cite{reddy06} have measured the [$\alpha$/Fe] abundance ratio for this object, which is 0.23, a value that at $\feh=-0.81$ appears too high for a low-$\alpha$ halo star, but it is only marginally consistent with the thick-disk trend, so it is not at all clear what $\alpha$-element population this star belongs to. Note that \cite{reddy06} and NS10 $\alpha$-element abundances are not necessarily on the same scale, so systematic differences between these two studies could be responsible for this apparent discrepancy. Combining all three data sets, we find that the high-$\alpha$ population shows [O/Fe] abundance ratios that decrease slightly from $\ofe\simeq0.60$ at $\feh=-2.3$ to $\ofe\simeq0.55$ at $-0.7$, and from there more abruptly to $\ofe\simeq0.35$ at $\feh\simeq-0.1$.

\begin{figure}
\centering
\includegraphics[bb=80 370 390 915,width=8.9cm]
{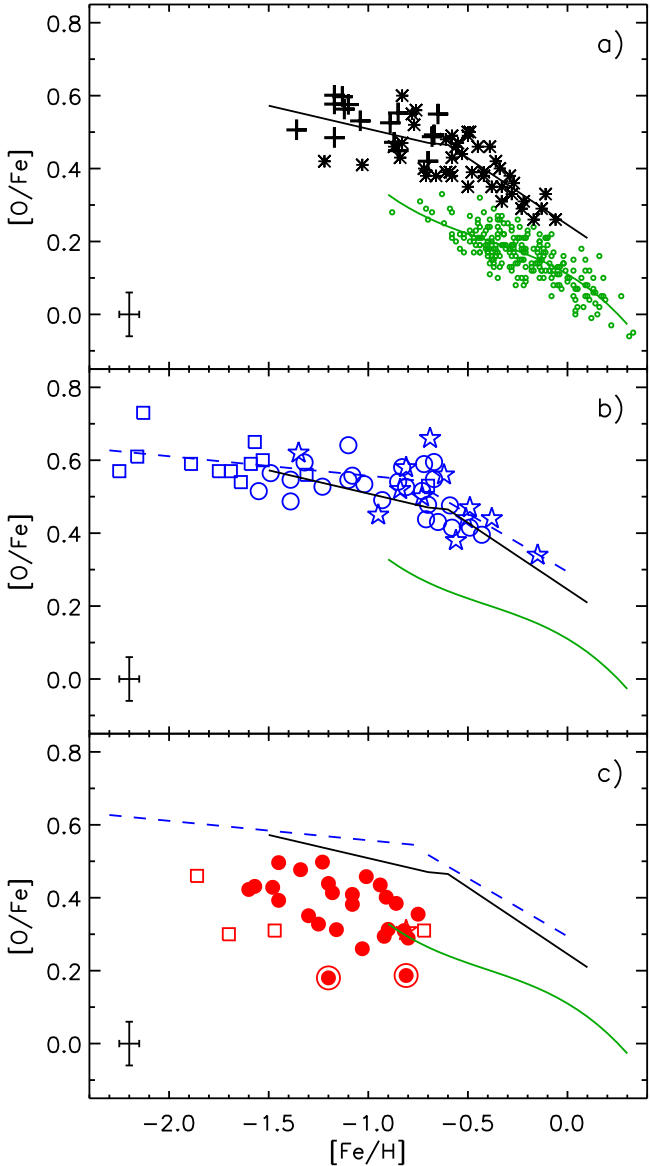}
\caption{{\bf a)} Oxygen abundance trends of thin- (circles) and thick-disk (crosses and asterisks) stars. The solid curve and broken line are fits to the thin- and thick-disk data, respectively, and they are shown also in panels b) and c). {\bf b)} Oxygen abundance trend of high-$\alpha$ halo stars. The dashed broken line is a fit to the high-$\alpha$ halo data and it is shown also in panel c). {\bf c)} Oxygen abundance pattern of low-$\alpha$ halo stars. Typical error bars are shown at the bottom left corner of each panel.}
\label{f:ofe_all_multi}
\end{figure}

The chemically-tagged stellar populations from Figure~\ref{f:ofe_all} have been re-plotted in Figure~\ref{f:ofe_all_multi}, in three panels: a) disk, b) high-$\alpha$, high-oxygen halo, and c) low-$\alpha$, low oxygen halo. In panel a), the solid line which has a knee at $\feh\simeq-0.6$ is a fit to the high-$\alpha$ disk data (the ``thick-disk'') whereas the solid curve is a cubic fit to the low-$\alpha$ disk data (the ``thin-disk''). These fits are reproduced in panels b) and c). In panel b), the dashed line which has a knee at $\feh\simeq-0.7$ is a fit to the high-$\alpha$ halo data, and is reproduced in panel c). The ``broken'' linear fits were made by allowing the knee location to be a free parameter. We monitored the 1\,$\sigma$ star-to-star scatter around these linear fits to make sure that these relations closely correspond to a minimum standard deviation. We also ensured that the fits resulted in two linear segments connecting smoothly at the knee.

The cubic fit to the thin-disk data has a star-to-star [O/Fe] scatter of only 0.04\,dex, which is compatible with a zero cosmic scatter, i.e., it suggests that observational errors alone explain it. This probably reflects the fact that the thin-disk stars have been born from well-mixed material at late stages in the history of Galactic evolution. The steep decline in [O/Fe] abundance ratio with increasing [Fe/H] has often been attributed to the chemical pollution of the interstellar medium by Type Ia supernovae (SN\,Ia), which dominate GCE only after a few billion years since the birth of the Galaxy.

The high-$\alpha$ halo and thick-disk populations appear very similar, but the linear fits suggest a small downwards offset for the thick disk relative to the halo. Also, the location of the knee seems to be about 0.1\,dex lower for the high-$\alpha$ halo group. The star-to-star scatter of these fits is about 0.06\,dex. Although observational errors are larger for these groups of stars, mainly because they are more distant and their atmospheric parameters cannot be determined as precisely as those of most thin-disk stars, some of this scatter could be real. Estimates of [O/Fe] errors are about 0.05\,dex for most of these objects. We note also that the star-to-star scatter of the thick-disk stars with $\feh<-0.6$ is about 0.07\,dex, suggesting that a less well-mixed gas, or more likely a mixture of different gases, gave origin to these objects compared to the high-$\alpha$ halo stars.

The fact that the high-$\alpha$ halo and thick-disk star oxygen abundance trends are best fit with a broken line suggests that their more metal-poor members were born from gas enriched mainly by the pollution of massive stars, whose yields (at the end of their lives as Type\,II supernovae, SNII) have large O/Fe abundance ratios. The shallow negative slope of the [O/Fe]--[Fe/H] relation at low $\feh$ ($\lesssim-0.65$) could be fully explained by metallicity-dependent SNII yields (R07). Later ($\feh\gtrsim-0.6$), SN\,Ia contributed significantly to the chemical evolution of these populations, quickly lowering their [O/Fe] ratios while $\feh$ continued to increase. That the high-$\alpha$ halo stars have a knee at a slightly lower $\feh$ is a possible indication that the transition happened earlier in the halo than in the thick disk, or that the star formation rate was somewhat slower in the halo. Note that the $\feh$ errors are about 0.05\,dex in this compilation of oxygen abundances, while the difference in $\feh$ for the location of the high-$\alpha$ halo and thick-disk knees is about 0.1\,dex, implying that this difference, although marginally consistent with zero within the errors, is unlikely due to observational uncertainties.

As shown in Figure~\ref{f:ofe_all_multi}c, the low-$\alpha$ halo stars do not follow any of the other [O/Fe] trends. They appear as a continuation of the thin disk, but this connection is very weak on light of other well-known observational evidence such as the stellar kinematics and age (thin-disk stars are younger and have cold kinematics, the complete opposite of what is normal in halo stars). On average, and excluding the stars with very low oxygen abundances discussed above, these stars have [O/Fe] abundance ratios that are about 0.2\,dex lower than those of high-$\alpha$ and thick-disk stars. There seems to be a downwards trend of [O/Fe] with increasing [Fe/H], but the star-to-star scatter is large. In any case, the mean [O/Fe] in the low-$\alpha$ halo population decreases from $\ofe\simeq0.5$ at $\feh\simeq-1.8$ to $\ofe\simeq0.35$ at $\feh\simeq-0.7$.

Low [O/Fe] abundance ratios at low $\feh$ have been typically attributed to populations with a history of slow star formation rate. They have been observed in the dwarf satellite galaxies of the Milky Way, suggesting that the low-$\alpha$, low-oxygen halo stars are remnants of merger processes that occurred early in the history of our Galaxy, or perhaps stars that have been stripped from their parent satellite galaxies as they came close or passed through the solar neighborhood in their Galactic orbits.

\subsection{Kinematics and Chemical Abundances} \label{s:kinematics}

\begin{figure}
\includegraphics[bb=75 368 450 630,width=9.0cm]
{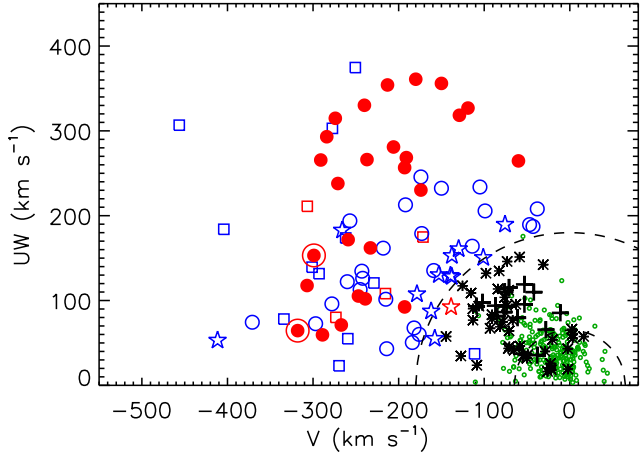}
\caption{Toomre diagram for the stars plotted in Figure~\ref{f:ofe_all}. Dashed lines correspond to speeds of 65 and 180\,\kms.}
\label{f:toomre}
\end{figure}

A very important tool to explore further the GCE interpretations is the Toomre diagram, which is shown for our sample of stars in Figure~\ref{f:toomre}. Here we plot the Galactic space velocity $V$ (which goes in the direction of disk rotation and is measured relative to the Sun), and the other two components combined: $UW=\sqrt{U^2+W^2}$ ($U$ is defined as the Galactic space velocity component towards the Galactic center while $W$ is the component perpendicular to the plane). This diagram is sometimes used to separate stars kinematically into thin-disk, thick-disk, and halo members. Although such kinematic membership criteria have been proven useful, particularly for solar neighborhood stars, as larger samples of stars are analyzed using better spectroscopic data and performing more careful elemental abundance work, we are starting to see that kinematics and chemical composition do not make a one-to-one correspondence for the local stellar populations. For example, although thin-disk stars tend to have cooler kinematics (i.e., $V$ closer to zero and low $UW$) compared to thick-disk members (lagging $V$ velocities and higher $UW$), there is an important number of stars with chemical composition typical of that of a thin-disk (thick-disk) star which have thick-disk (thin-disk) kinematics. The observed fraction of these stars with ``ambiguous'' kinematics and chemical abundances cannot be fully accounted for by observational errors (Ram\'irez, Allende Prieto, \& Lambert, in preparation).

Figure~\ref{f:toomre} shows that the chemically-tagged thin-disk stars, as a group, tend to have cold kinematics, i.e., they rotate fast ($V$ closer to zero) and do not depart much (small $UW$) from the mid-plane of the Milky Way's disk. Thick-disk stars, on the other hand, rotate slower (i.e., lag the Sun and the thin-disk stars, as a group) and have larger $U$ and $W$ velocities. This has been known for many years \cite[e.g.,][]{soubiran93,soubiran03}, and it has been used to select one group of stars or the other in chemical abundance studies \cite[e.g.,][]{bensby04,reddy06}. However, there is significant overlap between the two groups, and because of that, kinematic selection of thin/thick disk stars has often avoided the intermediate region, leading to thin/thick disk abundance patterns that may be affected by severe kinematic biases. An extended discussion of this possible bias will be given in Ram\'irez, Allende Prieto, and Lambert (in preparation). Here, we note that, although the thin-disk and thick-disk groups appear on average to separate kinematically, there is a significant number of stars with thin-disk abundances but thick-disk kinematics, and vice versa. If we make a simple kinematic separation, with a boundary of total speed equal to 65\,\kms, which corresponds to the inner dashed line shown in Figure~\ref{f:toomre}, we find that these stars with ambiguous kinematics and elemental abundances amount to about 30\,\%\ of each sample, a number that is certainly not negligible, as has been sometimes assumed or ignored altogether.

The thick disk has long been thought to be formed from a violent merger event early in the history of the Galaxy \cite[e.g.,][]{quinn93}. This merger would have destroyed a previously formed disk, heating the stars into more eccentric orbits. It is suggested by some authors that thick-disk stars could have been formed in the merging galaxies as well as in the original disk \cite[e.g.,][]{abadi03,brook05,kobayashi11}. These scenarios, however, are not able to fully explain the large fraction of stars with ambiguous kinematics and abundances. An alternative picture, revived by the work of \cite{sellwood02}, involves internal processes, particularly radial mixing, which explains many of the solar neighborhood observables \cite[e.g.,][]{schonrich09a,schonrich09,loebman11}. Nevertheless, recent SEGUE/SDSS observations of the Galaxy on a larger scale appear to be inconsistent with these models \cite[e.g.,][see, however, \citealt{bovy12_nothick,bovy12_vertical,bovy12_spatial}]{schlesinger12}. The issues of sample completeness and sample selection functions need to be fully addressed before observations of solar neighborhood stars are used to determine which scenario is more realistic. Upcoming and ongoing high-resolution spectroscopic surveys such as HERMES \citep{freeman10} and APOGEE \citep{allende08:apogee,majewski10} should allow us to solve this problem within the next decade.

Regarding the halo, there is not a one-to-one correspondence in chemical abundances and kinematics of high-$\alpha$ and low-$\alpha$ stars either, but an average trend was detected by NS10. They found that the low-$\alpha$ stars tend to have very low $V$ velocities, in fact most of them are in retrograde Galactic orbits. Four of the five low-$\alpha$ stars we found in M06 and R07 and which are not included in NS10 also have retrograde orbits, strengthening their conclusion. In their update of the R07 work, Ram\'irez, Allende Prieto, and Lambert (in preparation) also find a number of additional halo stars in retrograde orbits which have low oxygen abundances relative to the ``normal'' halo.

\begin{figure}
\includegraphics[bb=65 360 510 975,width=9.0cm]
{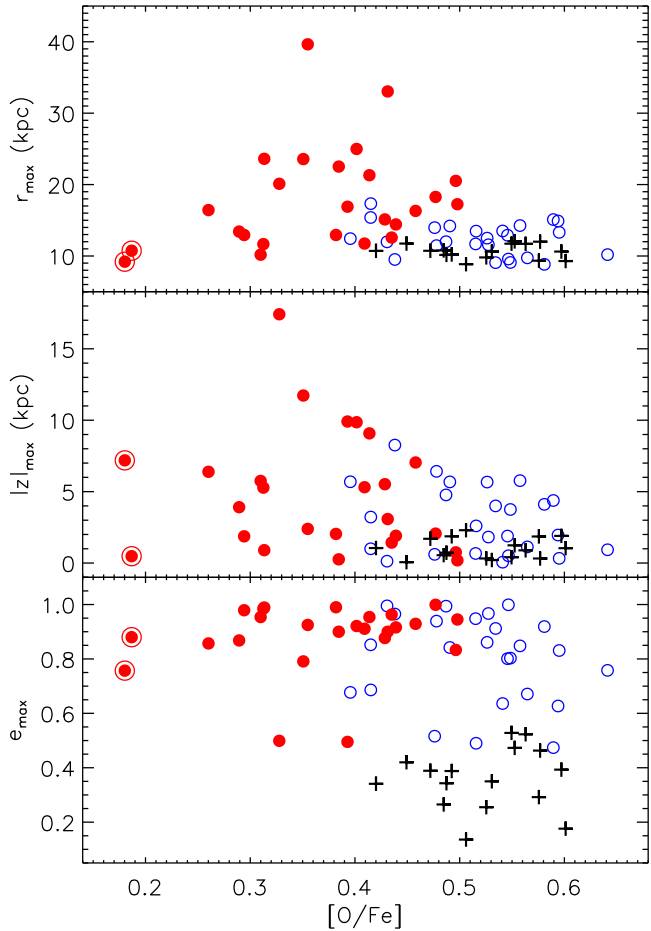}
\caption{Orbital parameters $r_\mathrm{max}$ (maximum horizontal distance from the Galactic center), $z_\mathrm{max}$ (maximum vertical distance from the Galactic plane), and $e_\mathrm{max}$ (maximum eccentricity) as a function of [O/Fe] for the NS10 stars analyzed in this work. Symbols are the same as in Figure~\ref{f:ofe}.}
\label{f:ofe_maxs}
\end{figure}

In Figure~\ref{f:ofe_maxs}, our oxygen abundances are plotted against the orbital parameters $r_\mathrm{max}$ (maximum distance from the Galactic center), $|z|_\mathrm{max}$ (maximum height with respect to the Galactic plane), and $e_\mathrm{max}$ (maximum eccentricity), as derived by \cite{schuster12} from 5\,Gyr orbit integrations computed using the NS10 star's kinematic data and a detailed semi-analytic model for the Milky Way potential. The orbital parameters plotted in Figure~\ref{f:ofe_maxs} correspond to those obtained by \cite{schuster12} with a realistic non-symmetrical Galaxy model. Note that the stars plotted in Figure~\ref{f:ofe_maxs} are those from Figure~\ref{f:ofe} and that the symbols used there are the same as those used in Figure~\ref{f:ofe} as well.

All of the high-$\alpha$, high-oxygen halo stars have orbits that do not go beyond 20\,kpc of distance from the Galactic center, contrary to about half of the low-$\alpha$, low-oxygen stars, which have $r_\mathrm{max}$ up to about 40\,kpc. Similarly, the high-$\alpha$, high-oxygen stars reach heights above the Galactic plane up to about 8\,kpc whereas some low-$\alpha$ stars have orbits that go about twice as high (see also Figure 8 in \citealt{schuster12}). The orbits of thick-disk members, on the other hand, are smaller in both directions, and they are also less eccentric than those of both types of halo stars. In fact, a majority of the latter have $e_\mathrm{max}$ greater than 0.8, whereas thick-disk stars have $e_\mathrm{max}\simeq0.35\pm0.15$. Note also that there are more high-$\alpha$, high-oxygen stars with $e_\mathrm{max}$ lower than 0.8 than low-$\alpha$, low-oxygen stars. The implications of these distinct orbital distributions were already discussed by \cite{schuster12}. In particular, they reinforce the idea that the two halo populations require different formation scenarios, with the low-$\alpha$ group being accreted stars. Interestingly, the distribution of $r_\mathrm{max}$ values for globular clusters is fully consistent with the scatter of $r_\mathrm{max}$ values seen in Figure~\ref{f:ofe_maxs} for the low-$\alpha$, low-oxygen halo stars \citep[e.g.,][]{dauphole96,dinescu99}.

\subsection{Field Halo Stars Born in Globular Clusters} \label{s:gc}

Based on their kinematics, NS10 proposed that the low-$\alpha$ halo stars could have been born in the dwarf satellite galaxies of the Milky Way, with some of them probably originating from the globular cluster (GC) $\omega$\,Cen. The more detailed chemical composition analysis made by these authors in \cite{nissen11}, however, revealed more differences (e.g., in $\alpha$, Na, and Ba/Y) than similarities (e.g., in Ni and Cu) between the low-$\alpha$ halo stars and $\omega$\,Cen. As suggested by them, perhaps chemical evolution in $\omega$\,Cen was different for its inner (or more bound) and outer (less bound) regions, explaining the present-day differences.

The association of groups of field halo stars with $\omega$\,Cen based on chemical analysis is tempting, as the many examples that can be found in the literature demonstrate, including the NS10 work. We must be reminded, however, that of all of the Milky Way's GCs, $\omega$\,Cen is the most complex example, exhibiting a wide range of stellar ages \cite[e.g.,][]{hughes00,stanford06} and metallicities \cite[e.g.,][]{norris95_wCen,frinchaboy02}. The latter imply that chemical evolution within the cluster has occurred following not a single but a number of episodes of star formation. Indeed, large chemical abundance surveys of $\omega$\,Cen stars suggest distinct chemical evolution paths followed by a number of clearly identified sub-populations \cite[e.g.,][]{johnson10_wCen,marino11}.

One of the most notable chemical properties of $\omega$\,Cen, observed also in most other GCs, is the so-called Na-O anti-correlation \cite[e.g.,][]{norris95,gratton01,gratton07,carretta09,johnson10_wCen,dantona11,marino11}, a property that is not seen in field halo stars. It has been suggested that this anti-correlation is due to {\it in-situ} mixing of intermediate-mass AGB star nucleosynthesis products \cite[e.g.,][]{ventura01,gratton04}. With our oxygen abundance data and the Na abundances from NS10 we can now explore another possible connection in the form of the Na-O anti-correlation.

\begin{figure}
\includegraphics[bb=70 372 450 626,width=8.8cm]
{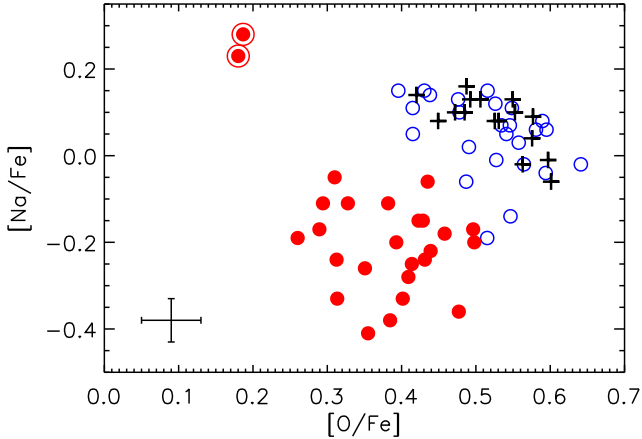}
\caption{[Na/Fe] versus [O/Fe] relation for the stars in Figure~\ref{f:ofe}. Sodium abundances are from \cite{nissen10}. Typical error bars are shown at the bottom left corner.}
\label{f:nao}
\end{figure}

In Figure~\ref{f:nao} we plot [Na/Fe] versus [O/Fe] for the stars studied in this work. With the exception of G53-41 and G150-40, the low-$\alpha$, low-oxygen abundance halo stars all have sub-solar [Na/Fe] abundance ratios. This observation is consistent with the oxygen and sodium abundance data for $\omega$\,Cen red giants by \cite{norris95}. A similar conclusion can be reached by looking at the \cite{johnson10_wCen} or \cite{marino11} larger data sets. We note, however, that the spread of oxygen and sodium abundances measured in $\omega$\,Cen stars is large, and in fact the [Na/Fe] versus [O/Fe] relation in $\omega$\,Cen overlaps also with the location of the high-$\alpha$ halo and thick-disk stars studied in this work. Only the presence of G53-41 and G150-40, the stars with lowest [O/Fe] and highest [Na/Fe] in our sample, hints at an Na-O anti-correlation for the low-$\alpha$ group. In fact, their peculiar chemical composition can be attributed to pollution by nearby AGB stars to the proto-stellar gas, in a similar fashion to globular cluster stars (NS10), although note that the abundance anomalies could also be due to fast rotating massive stars \cite[e.g.,][]{decressin07}. In any case, it is clear that the majority of low-$\alpha$ stars do not exhibit an obvious Na-O anti-correlation, which in principle further weakens the $\omega$\,Cen connection. Nevertheless, it is important to point out that there are no stars with this type of peculiar composition in the high-$\alpha$ halo or thick-disk groups.

The stars G53-41 and G150-40 are the only field stars showing the classical signatures of abundance anomalies in second generation GC stars, i.e., enhanced Na and depleted O. Other abundance peculiarities are also discussed in Section~\ref{s:other}. Since we have analyzed 67 stars, we conclude that the fraction of metal-poor field stars originating from second-generation GC stars is about 3\,\%. Adopting a binomial distribution, which is appropriate in this case given the relatively low number of objects and the fact that there are two possible ``outcomes'' for each star, i.e., field and GC, an error bar can be estimated from the variance of the probability distribution \cite[e.g.,][Chapter 3]{bevington69}: $\sigma^2=np(1-p)$, where $n=67$ is the number of stars and $p$ the probability of ``success'' ($p=2/67=0.03$). We find $\sigma=1.4$, which implies a probability error of $1.4/67=2$\,\%.

Of course, the actual fraction of halo field stars originally formed in GCs may be significantly higher than the value of $3\pm2$\,\% derived above, as the clusters have likely also contributed to the halo field with ``normal'' (i.e., first generation) stars, which may be hard to distinguish from the bulk of halo field stars observed today. Thus, our oxygen abundances and the Na abundances from NS10 suggest that the fraction of halo stars born in globular clusters is {\it at least} $3\pm2$\,\%.\footnote{Errors in our [O/Fe] and NS10's [Na/Fe] abundance ratios are too small to have a significant impact on this lower limit.} Indeed, although the fraction of field metal-poor giants with anomalous CN and CH bands (typical of second-generation GC stars) is only 3\,\%  according to \cite{martell11}, a minimum of 17\,\% of the present-day mass in the halo field originated from GCs. Using the binomial distribution on the \cite{martell11} results, we obtain a more precise value of $2.85\pm0.70$\,\%\ for the first of these fractions, in excellent agreement with our estimate of $3\pm2$\,\%. Note, however, that our highly reliable O and Na abundances connect these objects to GCs in a more direct way than the intensities of CN and CH bands.

\begin{figure*}
\includegraphics[bb=70 360 450 941,width=9.1cm]
{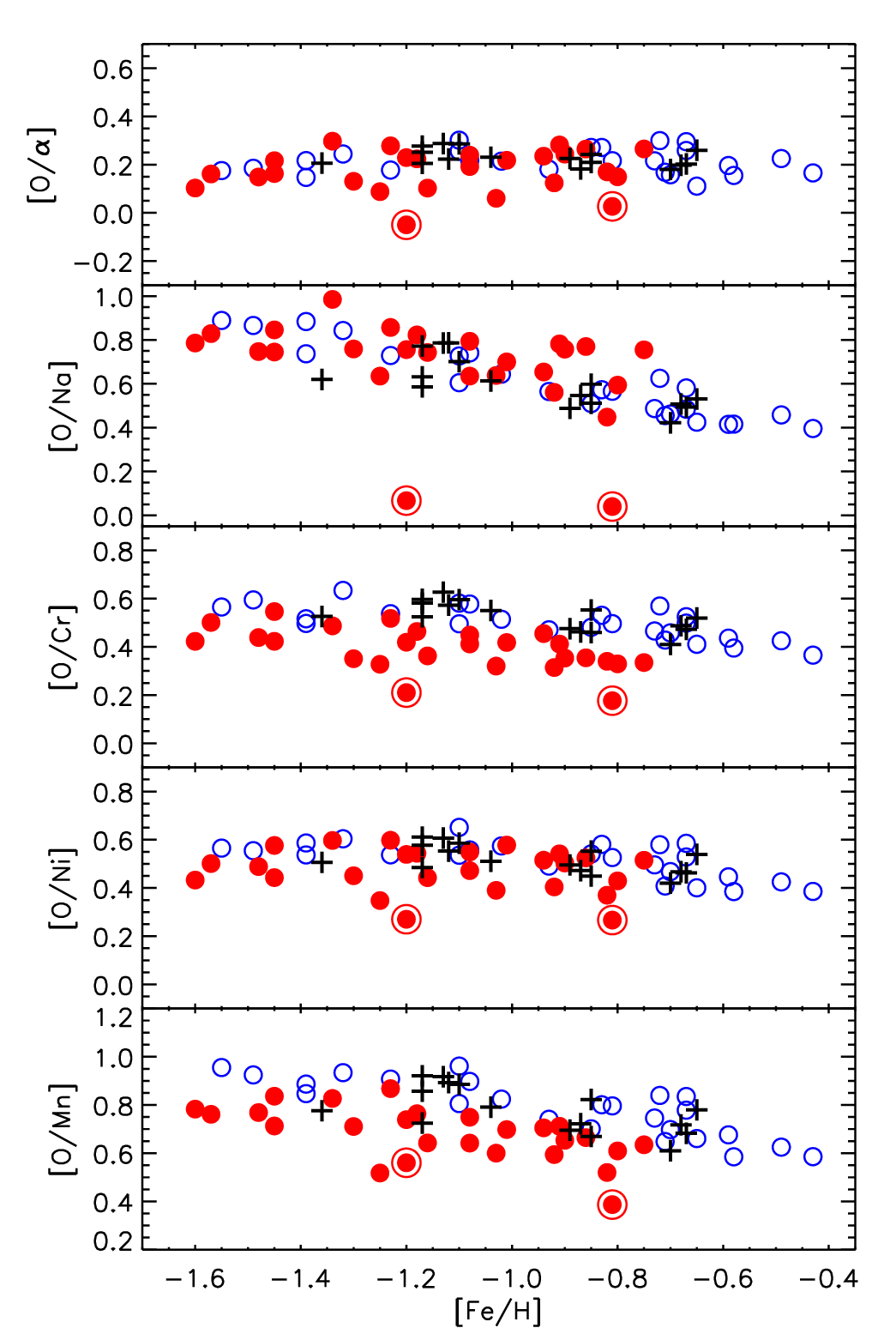}
\includegraphics[bb=70 360 450 941,width=9.1cm]
{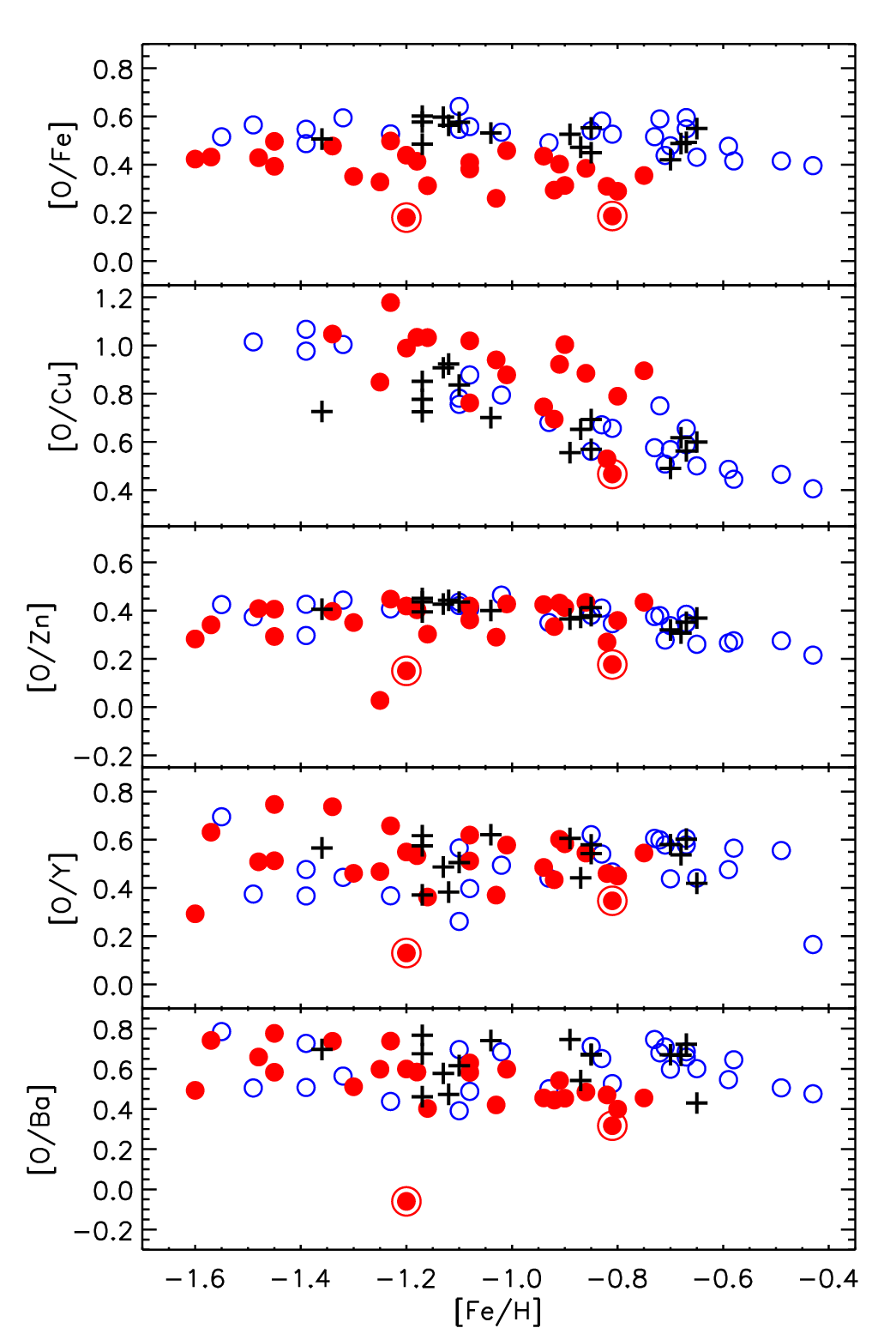}
\caption{[O/X] abundance ratios as a function of [Fe/H] for the stars in Figure~\ref{f:ofe}. Elemental abundances other than that of oxygen are from \cite{nissen10}. Note that the [O/X] axis range is 1.0\,dex in all panels except [O/Ba].}
\label{f:ox_feh}
\end{figure*}

A comparison between the two peculiar low-$\alpha$ stars (G53-41 and G150-40) and the Na and O abundances observed in second generation stars in the globular cluster M71 is instructive, as this cluster has $\feh=-0.8$ \citep{melendez09:m71}. In relation to first generation stars, the second generation stars in M71 have O depleted by only about 0.1\,dex while Na is enhanced by about 0.4\,dex \citep{melendez09:m71}. Similarly, for M4 ($\feh=-1.1$), \cite{marino11_m4} find an O depletion of about 0.25\,dex and a Na enhancement of about 0.4\,dex. These abundance variations are fully compatible with the ones observed between G53-41 and G150-40 and the bulk of low-$\alpha$ halo stars (Figure~\ref{f:nao}). While oxygen in these two dwarfs is depleted by about 0.2$\pm$0.1 dex, Na seems enhanced by about 0.4$\pm$0.1 dex. This quantitative comparison reinforces the idea that the two anomalous dwarfs may have been formed in GCs. If that is the case, G53-41 and G150-40 would be the first field stars with firm O/Na signatures of being originated in GCs. The discovery of these signatures certainly shows the benefit of high precision differential abundance studies.

\subsection{Oxygen Compared to Other Elements} \label{s:other}

It is interesting to note that, even though the high-$\alpha$, low-$\alpha$, and thick-disk stars studied in this work separate in both [Na/Fe] (Figure~6 in NS10) and [O/Fe] (our Figure~\ref{f:ofe}) when each of these abundance ratios is plotted against [Fe/H], the run of [O/Na] with [Fe/H] is essentially indistinguishable between the three groups, as shown by Figure~\ref{f:ox_feh}. The obvious exceptions are, again, G53-41 and G150-40. This striking similarity in elemental abundance ratios [O/X] between the three groups is seen also for the $\alpha$-elements as well as Zn. Other elements show systematic offsets in the distribution of [O/X] abundance ratios of the three stellar populations, although for Y and Ba the star-to-star scatter is too large to distinguish them, if present.

In Figure~\ref{f:ox_feh}, the range in the [O/X] axis has been set equal to 1.0\,dex for all panels, except that for Ba, for which it is 1.2\,dex. This allows a fair comparison of the star-to-star scatter differences and offsets between high-$\alpha$ halo stars, low-$\alpha$ halo stars, and thick-disk stars. For example, it is clear that the run of [O/Zn] with [Fe/H] is the tightest of all, especially if we exclude the three low-$\alpha$ stars with the lowest [O/Zn]. Two of these three stars are G53-41 and G150-40, which have been mentioned before as having peculiar composition. The other object, which is the one with the lowest [O/Zn] in our sample, is G112-43. As pointed out by \cite{nissen11}, this object is one of the components of a wide binary, and, along with its companion (with which it has consistent chemical abundances), they seem to also have peculiar abundances (in [Mn/Fe] and [Zn/Fe]) when compared to the mean low-$\alpha$ trends.

Excluding G53-41, G150-40, and G112-43 from Figure~\ref{f:ox_feh} results in tighter correlations and stronger population similarities for [O/$\alpha$], [O/Na], and [O/Zn]. In addition to the latter abundance ratios, the outlier stars appear to be very peculiar in [O/Y] and [O/Ba]. The most extreme case is [O/Na], but, interestingly, only for G53-41 and G150-40, i.e., not for the binary component G112-43. For [O/Y] and [O/Ba], only one of the two stars with very low oxygen abundance (G53-41) has an unusually high (relative to other low-$\alpha$ stars) [Ba/Fe] abundance ratio ($\mathrm{[Ba/Fe]}=+0.24$ according to NS10), whereas the other (G150-40) shows marginally low [O/Y] and [O/Ba] which may be considered normal for a low-$\alpha$ halo star. Barium abundance anomalies relative to normal field halo stars have also been seen in $\omega$\,Cen members \cite[e.g.,][]{norris95,smith00}, but the enhancement observed in those stars is significantly higher than that measured in G53-41.

\section{Conclusions}

Non-LTE oxygen abundances from the 777\,nm \oi\ triplet lines have been derived for as many as possible of the stars in the work by NS10. These authors have derived very precise atmospheric parameters and elemental abundances (excluding oxygen) for their sample stars, allowing them to clearly separate the field halo stars into low-$\alpha$ and high-$\alpha$ groups.

We find the run of [O/Fe] abundance ratios with $\feh$ of high-$\alpha$ halo and thick-disk stars very similar, while that of low-$\alpha$ halo stars is systematically lower by about 0.2\,dex and it has, in general, a larger star-to-star scatter compared to the other two groups. A few additional low-$\alpha$, low-oxygen abundance halo stars are identified in previously published works. Their kinematic properties strengthen the hypothesis by NS10 that these objects may have originated in dwarf satellite galaxies early in the history of the Milky Way. A connection between the low-$\alpha$, low oxygen halo stars and $\omega$\,Cen is not well established, unless assumptions about the early chemical abundance distribution within this extremely complex globular cluster are made.

Our oxygen abundance data for the three groups of stars studied by NS10 exhibit a behavior that is similar to that of the $\alpha$-elements. The exceptions are two stars, G53-41 and G150-40, which seem to be the first firm candidates of field halo stars born in globular clusters, although probably not $\omega$\,Cen, which has been previously argued as one of the main contributors of low-$\alpha$ field halo stars. Both G53-41 and G150-40 show the classic signatures of abundance anomalies in globular cluster stars, namely very low oxygen and highly enhanced sodium abundances. Since these properties are seen in 2 of the 67 stars studied in this work, we estimate that the contribution of globular clusters to the local field metal-poor ($-1.6<\feh<-0.4$) stellar population is at least $3\pm2$\,\%.

\acknowledgments

We thank Dr.\ A.\ F.\ Marino for her helpful advice on the chemical properties of globular clusters. I.R.'s work was performed under contract with the California Institute of Technology (Caltech) funded by NASA through the Sagan Fellowship Program. J.M. would like to acknowledge support from USP (Novos Docentes), FAPESP (2010/17510-3) and CNPq (Bolsa de Produtividade). Work by J.C. has been supported by the Centro de Astrof\'isica FONDAP 15010003, the Centro de Astrof\'isica y Tecnolog\'ias Afines CATA del Proyecto Financiamiento Basal PFB06, and by NASA through Hubble Fellowship grant HST-HF-51239.01-A, awarded by the Space Telescope Science Institute, which is operated by the Association of Universities for Research in Astronomy, Inc., for NASA, under contract NAS5-26555.


\begin{deluxetable}{lrrrrrrrcc}
\tablecolumns{10}
\tablewidth{0pc}
\tablecaption{Stellar Parameters and Oxygen Abundances.\tablenotemark{1}}
\tabletypesize{\footnotesize}
\tablehead{\colhead{Star} & \colhead{$\teff$ (K)} & \colhead{$\logg$} & \colhead{[Fe/H]} & \colhead{[O/H]$_\mathrm{LTE}$} & \colhead{[O/Fe]$_\mathrm{LTE}$} & \colhead{[O/H]$_\mathrm{NLTE}$} & \colhead{[O/Fe]$_\mathrm{NLTE}$} & \colhead{$\alpha$/Fe} & \colhead{Source}}
\startdata
         g05-36 & 6013 & 4.23 & $-1.23$ & $-0.51\pm0.04$ & $ 0.72\pm0.05$ & $-0.70\pm0.03$ & $ 0.53\pm0.05$ &  high &    R07/McD-Tull \\
        g159-50 & 5624 & 4.37 & $-0.93$ & $-0.35\pm0.04$ & $ 0.58\pm0.05$ & $-0.44\pm0.04$ & $ 0.49\pm0.05$ &  high &    R07/McD-Tull \\
        g170-56 & 5994 & 4.12 & $-0.92$ & $-0.47\pm0.04$ & $ 0.45\pm0.05$ & $-0.63\pm0.03$ & $ 0.29\pm0.04$ &   low &    R07/McD-Tull \\
        g176-53 & 5523 & 4.48 & $-1.34$ & $-0.72\pm0.05$ & $ 0.62\pm0.06$ & $-0.86\pm0.07$ & $ 0.48\pm0.08$ &   low &    R07/McD-Tull \\
         g18-28 & 5372 & 4.41 & $-0.83$ & $-0.20\pm0.04$ & $ 0.63\pm0.05$ & $-0.25\pm0.03$ & $ 0.58\pm0.05$ &  high &    R07/McD-Tull \\
        g180-24 & 6004 & 4.21 & $-1.39$ & $-0.65\pm0.04$ & $ 0.74\pm0.05$ & $-0.84\pm0.04$ & $ 0.55\pm0.05$ &  high &    R07/McD-Tull \\
        g188-22 & 5974 & 4.18 & $-1.32$ & $-0.52\pm0.03$ & $ 0.80\pm0.04$ & $-0.73\pm0.03$ & $ 0.59\pm0.04$ &  high &    R07/McD-Tull \\
         g56-36 & 5933 & 4.28 & $-0.94$ & $-0.35\pm0.03$ & $ 0.59\pm0.04$ & $-0.50\pm0.03$ & $ 0.44\pm0.04$ &   low &    R07/McD-Tull \\
         g85-13 & 5628 & 4.38 & $-0.59$ & $-0.05\pm0.04$ & $ 0.54\pm0.05$ & $-0.11\pm0.04$ & $ 0.48\pm0.05$ &  high &    R07/McD-Tull \\
         g99-21 & 5487 & 4.39 & $-0.67$ & $-0.08\pm0.04$ & $ 0.59\pm0.05$ & $-0.12\pm0.04$ & $ 0.55\pm0.05$ &  high &    R07/McD-Tull \\
       hd103723 & 5938 & 4.19 & $-0.80$ & $-0.38\pm0.03$ & $ 0.42\pm0.04$ & $-0.51\pm0.03$ & $ 0.29\pm0.04$ &   low &    R07/McD-Tull \\
       hd106516 & 6196 & 4.42 & $-0.68$ & $-0.01\pm0.04$ & $ 0.67\pm0.05$ & $-0.19\pm0.03$ & $ 0.49\pm0.04$ &   tdk &    R07/McD-Tull \\
       hd111980 & 5778 & 3.96 & $-1.08$ & $-0.31\pm0.04$ & $ 0.77\pm0.05$ & $-0.52\pm0.03$ & $ 0.56\pm0.04$ &  high &    R07/McD-Tull \\
      hd114762a & 5856 & 4.21 & $-0.70$ & $-0.14\pm0.03$ & $ 0.56\pm0.05$ & $-0.28\pm0.03$ & $ 0.42\pm0.04$ &   tdk &    R07/McD-Tull \\
       hd126681 & 5507 & 4.45 & $-1.17$ & $-0.46\pm0.05$ & $ 0.71\pm0.06$ & $-0.57\pm0.05$ & $ 0.60\pm0.06$ &   tdk &    R07/McD-Tull \\
       hd132475 & 5646 & 3.76 & $-1.49$ & $-0.64\pm0.04$ & $ 0.85\pm0.05$ & $-0.93\pm0.04$ & $ 0.56\pm0.05$ &  high &    R07/McD-Tull \\
       hd159482 & 5737 & 4.31 & $-0.73$ & $-0.09\pm0.03$ & $ 0.64\pm0.04$ & $-0.21\pm0.03$ & $ 0.52\pm0.04$ &  high &    R07/McD-Tull \\
       hd160693 & 5714 & 4.27 & $-0.49$ & $ 0.02\pm0.05$ & $ 0.51\pm0.06$ & $-0.07\pm0.04$ & $ 0.42\pm0.05$ &  high &    R07/McD-Tull \\
       hd163810 & 5501 & 4.56 & $-1.20$ & $-0.67\pm0.06$ & $ 0.53\pm0.07$ & $-0.76\pm0.04$ & $ 0.44\pm0.05$ &   low &    R07/McD-Tull \\
        hd17820 & 5773 & 4.22 & $-0.67$ & $-0.05\pm0.05$ & $ 0.62\pm0.06$ & $-0.18\pm0.04$ & $ 0.49\pm0.05$ &   tdk &    R07/McD-Tull \\
       hd179626 & 5853 & 4.16 & $-1.02$ & $-0.31\pm0.03$ & $ 0.71\pm0.04$ & $-0.49\pm0.03$ & $ 0.53\pm0.04$ &  high &    R07/McD-Tull \\
       hd189558 & 5620 & 3.81 & $-1.12$ & $-0.35\pm0.05$ & $ 0.77\pm0.06$ & $-0.56\pm0.04$ & $ 0.56\pm0.05$ &   tdk &    R07/McD-Tull \\
       hd219617 & 5862 & 4.28 & $-1.45$ & $-0.77\pm0.05$ & $ 0.68\pm0.06$ & $-0.95\pm0.04$ & $ 0.50\pm0.05$ &   low &    R07/McD-Tull \\
        hd22879 & 5759 & 4.25 & $-0.85$ & $-0.15\pm0.03$ & $ 0.70\pm0.04$ & $-0.30\pm0.03$ & $ 0.55\pm0.04$ &   tdk &     R07/HET-HRS \\
       hd241253 & 5831 & 4.31 & $-1.10$ & $-0.36\pm0.07$ & $ 0.74\pm0.07$ & $-0.52\pm0.06$ & $ 0.58\pm0.06$ &   tdk &    R07/McD-Tull \\
      hd250792a & 5489 & 4.47 & $-1.01$ & $-0.49\pm0.03$ & $ 0.52\pm0.05$ & $-0.55\pm0.04$ & $ 0.46\pm0.05$ &   low &    R07/McD-Tull \\
        hd76932 & 5877 & 4.13 & $-0.87$ & $-0.22\pm0.03$ & $ 0.65\pm0.04$ & $-0.40\pm0.03$ & $ 0.47\pm0.04$ &   tdk &    R07/VLT-UVES \\
         g20-15 & 6050 & 4.34 & $-1.45$ & $-0.91\pm0.06$ & $ 0.54\pm0.07$ & $-1.06\pm0.05$ & $ 0.39\pm0.06$ &   low &           HIRES \\
         g66-22 & 5236 & 4.41 & $-0.86$ & $-0.47\pm0.09$ & $ 0.39\pm0.09$ & $-0.48\pm0.09$ & $ 0.38\pm0.09$ &   low &           HIRES \\
         g82-05 & 5277 & 4.45 & $-0.75$ & $-0.41\pm0.04$ & $ 0.34\pm0.05$ & $-0.40\pm0.04$ & $ 0.35\pm0.05$ &   low &           HIRES \\
        hd51754 & 5767 & 4.29 & $-0.58$ & $-0.05\pm0.06$ & $ 0.53\pm0.07$ & $-0.17\pm0.05$ & $ 0.41\pm0.05$ &  high &           HIRES \\
       hd175179 & 5713 & 4.33 & $-0.65$ & $ 0.01\pm0.09$ & $ 0.66\pm0.10$ & $-0.10\pm0.07$ & $ 0.55\pm0.08$ &   tdk &           HIRES \\
       hd222766 & 5334 & 4.27 & $-0.67$ & $-0.03\pm0.06$ & $ 0.64\pm0.07$ & $-0.07\pm0.05$ & $ 0.60\pm0.06$ &  high &           HIRES \\
         g15-23 & 5297 & 4.57 & $-1.10$ & $-0.39\pm0.05$ & $ 0.71\pm0.06$ & $-0.46\pm0.04$ & $ 0.64\pm0.05$ &  high &           HIRES \\
         g24-13 & 5673 & 4.31 & $-0.72$ & $-0.02\pm0.04$ & $ 0.70\pm0.05$ & $-0.13\pm0.04$ & $ 0.59\pm0.05$ &  high &           HIRES \\
       hd230409 & 5318 & 4.54 & $-0.85$ & $-0.29\pm0.05$ & $ 0.56\pm0.06$ & $-0.31\pm0.04$ & $ 0.54\pm0.05$ &  high &           HIRES \\
        g119-64 & 6181 & 4.18 & $-1.48$ & $-0.88\pm0.05$ & $ 0.60\pm0.06$ & $-1.05\pm0.04$ & $ 0.43\pm0.05$ &   low &           HIRES \\
       hd233511 & 6006 & 4.23 & $-1.55$ & $-0.85\pm0.06$ & $ 0.70\pm0.07$ & $-1.03\pm0.06$ & $ 0.52\pm0.07$ &  high &           HIRES \\
       hd148816 & 5832 & 4.14 & $-0.71$ & $-0.12\pm0.05$ & $ 0.59\pm0.06$ & $-0.27\pm0.03$ & $ 0.44\pm0.05$ &  high &      MIKE/Jul11 \\
       hd193901 & 5663 & 4.39 & $-1.08$ & $-0.57\pm0.03$ & $ 0.51\pm0.04$ & $-0.67\pm0.03$ & $ 0.41\pm0.04$ &   low &      MIKE/Jul11 \\
         hd3567 & 6051 & 4.02 & $-1.16$ & $-0.66\pm0.03$ & $ 0.50\pm0.04$ & $-0.85\pm0.03$ & $ 0.31\pm0.04$ &   low &      MIKE/Jul11 \\
      cd-61.282 & 5759 & 4.31 & $-1.23$ & $-0.57\pm0.05$ & $ 0.66\pm0.06$ & $-0.73\pm0.05$ & $ 0.50\pm0.06$ &   low &      MIKE/Jul11 \\
       hd121004 & 5669 & 4.37 & $-0.70$ & $-0.14\pm0.04$ & $ 0.56\pm0.05$ & $-0.22\pm0.03$ & $ 0.48\pm0.04$ &  high &      MIKE/Jul11 \\
       hd199289 & 5810 & 4.28 & $-1.04$ & $-0.35\pm0.03$ & $ 0.69\pm0.04$ & $-0.51\pm0.03$ & $ 0.53\pm0.04$ &   tdk &      MIKE/Jul11 \\
       hd194598 & 5934 & 4.33 & $-1.08$ & $-0.56\pm0.03$ & $ 0.52\pm0.04$ & $-0.70\pm0.03$ & $ 0.38\pm0.04$ &   low &      MIKE/Jul11 \\
       hd284248 & 6135 & 4.25 & $-1.57$ & $-0.98\pm0.04$ & $ 0.59\pm0.05$ & $-1.14\pm0.04$ & $ 0.43\pm0.05$ &   low &      MIKE/Sep11 \\
        hd59392 & 6012 & 3.91 & $-1.60$ & $-0.96\pm0.05$ & $ 0.64\pm0.06$ & $-1.18\pm0.05$ & $ 0.42\pm0.06$ &   low &      MIKE/Sep11 \\
     cd-33.3337 & 5979 & 3.86 & $-1.36$ & $-0.61\pm0.04$ & $ 0.75\pm0.05$ & $-0.85\pm0.04$ & $ 0.51\pm0.05$ &   tdk &      MIKE/Sep11 \\
     cd-57.1633 & 5873 & 4.28 & $-0.90$ & $-0.47\pm0.03$ & $ 0.43\pm0.04$ & $-0.59\pm0.03$ & $ 0.31\pm0.04$ &   low &      MIKE/Sep11 \\
       hd120559 & 5412 & 4.50 & $-0.89$ & $-0.32\pm0.04$ & $ 0.57\pm0.05$ & $-0.36\pm0.04$ & $ 0.53\pm0.05$ &   tdk &      MIKE/Sep11 \\
         g31-55 & 5638 & 4.30 & $-1.10$ & $-0.42\pm0.04$ & $ 0.68\pm0.05$ & $-0.55\pm0.05$ & $ 0.55\pm0.05$ &  high &      MIKE/Nov11 \\
         g05-19 & 5854 & 4.26 & $-1.18$ & $-0.61\pm0.04$ & $ 0.57\pm0.05$ & $-0.77\pm0.04$ & $ 0.41\pm0.05$ &   low &      MIKE/Nov11 \\
         g05-40 & 5795 & 4.17 & $-0.81$ & $-0.12\pm0.03$ & $ 0.69\pm0.04$ & $-0.28\pm0.03$ & $ 0.53\pm0.04$ &  high &      MIKE/Nov11 \\
        hd25704 & 5868 & 4.26 & $-0.85$ & $-0.26\pm0.07$ & $ 0.59\pm0.08$ & $-0.40\pm0.05$ & $ 0.45\pm0.06$ &   tdk &      MIKE/Nov11 \\
     cd-45.3283 & 5597 & 4.55 & $-0.91$ & $-0.46\pm0.03$ & $ 0.45\pm0.04$ & $-0.51\pm0.03$ & $ 0.40\pm0.05$ &   low &      MIKE/Nov11 \\
        g112-43 & 6074 & 4.03 & $-1.25$ & $-0.73\pm0.04$ & $ 0.52\pm0.05$ & $-0.92\pm0.04$ & $ 0.33\pm0.05$ &   low &      MIKE/Nov11 \\
     cd-51.4628 & 6153 & 4.31 & $-1.30$ & $-0.80\pm0.03$ & $ 0.50\pm0.05$ & $-0.95\pm0.03$ & $ 0.35\pm0.05$ &   low &      MIKE/Nov11 \\
        hd97320 & 6008 & 4.19 & $-1.17$ & $-0.48\pm0.05$ & $ 0.69\pm0.06$ & $-0.69\pm0.05$ & $ 0.48\pm0.05$ &   tdk &      MIKE/Nov11 \\
       hd205650 & 5698 & 4.32 & $-1.17$ & $-0.45\pm0.04$ & $ 0.72\pm0.05$ & $-0.59\pm0.04$ & $ 0.58\pm0.05$ &   tdk &      MIKE/Nov11 \\
     bd-21.3420 & 5808 & 4.26 & $-1.13$ & $-0.35\pm0.04$ & $ 0.78\pm0.05$ & $-0.53\pm0.04$ & $ 0.60\pm0.05$ &   tdk &      MIKE/Feb12 \\
     cd-43.6810 & 5945 & 4.26 & $-0.43$ & $ 0.12\pm0.04$ & $ 0.55\pm0.05$ & $-0.03\pm0.03$ & $ 0.40\pm0.04$ &  high &      MIKE/Feb12 \\
       hd113679 & 5672 & 3.99 & $-0.65$ & $-0.08\pm0.04$ & $ 0.57\pm0.05$ & $-0.22\pm0.03$ & $ 0.43\pm0.04$ &  high &      MIKE/Feb12 \\
        g150-40 & 5968 & 4.09 & $-0.81$ & $-0.49\pm0.05$ & $ 0.32\pm0.06$ & $-0.62\pm0.06$ & $ 0.19\pm0.06$ &   low &  McD-Tull/Apr12 \\
         g18-39 & 6040 & 4.21 & $-1.39$ & $-0.71\pm0.04$ & $ 0.68\pm0.05$ & $-0.90\pm0.04$ & $ 0.49\pm0.05$ &  high &       Akerman04 \\
       hd105004 & 5754 & 4.30 & $-0.82$ & $-0.42\pm0.03$ & $ 0.40\pm0.04$ & $-0.51\pm0.03$ & $ 0.31\pm0.04$ &   low &       Akerman04 \\
         g53-41 & 5859 & 4.27 & $-1.20$ & $-0.90\pm0.05$ & $ 0.30\pm0.06$ & $-1.02\pm0.05$ & $ 0.18\pm0.06$ &   low &       Akerman04 \\
         g75-31 & 6010 & 4.02 & $-1.03$ & $-0.58\pm0.04$ & $ 0.45\pm0.05$ & $-0.77\pm0.03$ & $ 0.26\pm0.04$ &   low &        Nissen02 \\

\enddata
\tablenotetext{1}{The atmospheric parameters $\teff,\logg,\feh$ and $\alpha$/Fe flags are from \cite{nissen10}.}
\label{t:ofe}
\end{deluxetable}

\end{document}